\pdfoutput=1
\documentclass[twocolumn,trackchanges]{aastex61}

\received{....}
\revised{.....}
\accepted{......}
\submitjournal{AJ}
\usepackage{txfonts}

\begin{document}

\title{Evolution of an asteroid family under YORP, Yarkovsky and collisions}

\author{Francesco Marzari}
\affiliation{Department of Physics and Astronomy, University of Padova,
via Marzolo 8, 35131, Padova, Italy}
\email{francesco.marzari@pd.infn.it}

\author{Alessandro Rossi}
\affiliation{IFAC-CNR, Via Madonna del Piano 10, 50019 Sesto Fiorentino, Italy}
\email{a.rossi@ifac.cnr.it}

\author{Oleksiy Golubov}
\affiliation{School of Physics and Technology, V.N. Karazin Kharkiv National University, 4 Svobody Sq., Kharkiv, 61022, Ukraine}
\email{leksiy.golubov@gmail.com}

\author{Daniel Scheeres}
\affiliation{Department of Aerospace Engineering Sciences, University of Colorado at Boulder, 429 UCB, Boulder, CO, 80309, USA}
\email{scheeres@colorado.edu}

\begin{abstract}
\noindent
 Any population of asteroids, like asteroid families, will disperse in semi--major axis due to the Yarkovsky effect. The amount of drift  is modulated by the asteroid spin state evolution which determines the balance between the diurnal and seasonal Yarkovsky force.  The asteroid's spin state is, in turn, controlled in part by the YORP effect. The otherwise smooth evolution of an asteroid can be abruptly altered by collisions, which can cause impulsive changes in the spin state  and can move the asteroid onto a different YORP track. In addition, collisions may also alter the YORP parameters by changing the superficial features and overall shape of the asteroid. Thus, the coupling between YORP and Yarkovsky is also strongly affected by the impact history of each body. To investigate this coupling we developed a statistical code modeling the time evolution of semi--major axis  under YORP--Yarkovsky coupling. It includes the contributions of NYORP (normal YORP), TYORP (tangential YORP) and collisions whose effects are deterministically calculated and not added in a statistical way. We find that both collisions and TYORP  increase the dispersion of a family in semi--major axis  by making the spin axis evolution less smooth and regular. We show that the evolution of a family's structure with time is complex  and collisions randomize the YORP evolution.  In our test families we do not observe the formation of a 'YORP--eye' in the semi--major axis vs. diameter distribution, even after a long period of time. If present, the 'YORP--eye' might be a relic of an initial ejection velocity pattern of the collisional fragments.

\end{abstract}


\section{Introduction}

 The disruption of large asteroids has occurred frequently over the age of the solar system, leading to the formation of asteroid families. The identification of these families and the dating of their formation epoch has been used to enrich our understanding of the evolution of the solar system. There are several approaches to the identification of asteroid families, although they all rely on identifying the collisional fragments through the clustering of asteroids in the space of proper elements \citep{zoran2002} via different methods such as the Hierarchical Clustering Method  (HCNM, \cite{zappala1990,zappala1992}) and the Wavelet Analysis Method  (WAM,  \cite{bendjoya1991}), possibly assisted by additional information concerning color and albedo \citep{masiero2013} of the putative family members. Another method to identify families is via  backward integration in time of their orbital elements, which should converge towards the orbit of the parent body. However, backwards integration works only for very young families with ages of some Myrs  \citep{nesvorny2002}.  

 The  family identification methods mentioned above are based on the assumption that proper elements, purged of secular perturbative terms,  are stable over long timescales comparable to the age of the solar system. However,  the Yarkovsky effect \citep{rubincam1987,rubincam95,vokru1999,bottke2001} (anisotropic emission of thermal radiation due to thermal inertia) can compromise this assumption, and may lead to a significant inward/outward radial drift for small asteroids at a rate that is inversely proportional to their size. 
Due to the Yarkovsky effect families disperse over time, with the halo of smaller family members expanding in the proper elements space. As a consequence, and depending on their age and size distribution, older families may become unrecognizable as statistically significant clusters with respect to the background population of asteroids, and thus would not be detected with the usual methods like HCM and WAM. 

To prevent this loss of information, \cite{spoto2015} and \cite{bolin2017} developed a more refined identification method based on the V--shape acquired by the family in the $a$ vs. $1/D$ plane due to the Yarkovsky dispersal, where  $a$ is the semi--major axis and $D$ the diameter of each body.
The width of this V-shape depends  on the Yarkovsky drift rate, which is determined by the physical properties of the family members like albedo, thermal inertia and rotation period and spin axis of
the asteroids \citep{vokru1999}. 
The YORP \citep{rubincam2000} effect plays a fundamental role in determining the temporal details of the V--shape since, as it can lead to short term variations of the obliquity of the asteroids which in turn affects the balance between the diurnal and seasonal Yarkovsky effect. YORP is a radiation torque due to scattered and thermally re-emitted 
sunlight and it is related to the overall shape of the asteroid and also to  middle- to small-scale irregularities of the surface. A complete review of the models developed to describe the spin axis evolution due to YORP is given in \cite{vokru2015}. 
In addition to these non-gravitational effects, collisions are also seen to be an important additional ingredient when predicting the evolution of the spin axis of an asteroid. While YORP acts to produce a slow continuous evolution in the spin axis and spin rate of an asteroid, collisions cause impulsive changes that depend on the impact geometry and energy. 

Once coupled with YORP and collisions, models of the Yarkovsky drift of family members have been successfully reproduced.   In \cite{vokru2006}  the structure in the semimajor axis--absolute magnitude plane of the Erigone asteroid family was successfully modeled. Here they accounted for the effects of the most energetic collisions by resetting the spin vector to a new random state, using a statistical model for the collisional evolution of asteroids. In \cite{bottke2015}  a stochastic form of the YORP effect was invoked to model the Eulaia family, related to changes in shape due to collisions and centrifugally driven reshaping. Noteworthy among such models are those of \cite{paolicchi2016} and \cite{paolicchi2019} which predicted that the coupled YORP--Yarkovsky evolution would produce a 'YORP--eye', a depletion of family members in the center of the V--shape based on the assumption of a YORP-driven clustering of asteroid spin axe close to either $0^o$ or $180^o$ of obliquity. 

 In this paper,  we use a different approach to follow the evolution with time in the $a$--$D$ plane of the putative members of a family. We adopt the recently developed model  of \cite{golubov2019} to compute the evolution of the spin vector of each asteroid due to radiation re--emission which incorporates both NYORP (Normal YORP due to the global shape of the body) and TYORP (Tangential YORP as in \cite{golukru2012,golubov2014,seve2015} due to small local features like boulders). In addition, we include the effects of collisions in a way that accounts not only for the large impact events but also for the more numerous small impacts whose cumulative effect can significantly interfere with the NYORP-TYORP evolution. The model is similar to that used in asteroid collisional evolution models and is described in detail in \cite{marzari2011Icar}. 
Due to the simple formulation of the NYORP-TYORP effect by 
\cite{golubov2019}, we can also reset the  model parameters due to collisions and reshaping when the breakup limit is approached. This combined model integrates state of the art models for YORP and collisions and their interactions to create a more accurate model for Yarkovsky migration of asteroid family members.  

We focus here on the theoretical predictions of this model for putative families, while in a forthcoming paper we will model some specific families. In Sect. \ref{YORP1} we  briefly summarize the NYORP+TYORP theory
and in Sect. \ref{MODEL} we  give the details of the numerical algorithm that includes NYORP-TYORP, collisions and Yarkovsky drift. Sect.\ref{TEST} is devoted to the description of the evolution of a small test family while Sect. \ref{TEST_BIG} outlines the evolution of a large putative family with parameters similar to those of the Koronis family. Finally, in Sect. \ref{CONCLU} we discuss our results and their implications.

\section{Evolution of the spin due to YORP} 
\label{YORP1}

The evolution of the rotation rate $\omega$ and obliquity $\varepsilon$ of each member of the family is computed exploiting the equations \citep{golubov2019}
\begin{eqnarray}
\label{eq_z}
I_z\frac{d\omega}{dt}=T_{z\,\mathrm{NYORP}}+T_{z\,\mathrm{TYORP}},\\
\label{eq_eps}
I_z\frac{d\varepsilon}{dt}=\frac{1}{\omega}T_{\varepsilon\,\mathrm{NYORP}}.
\end{eqnarray}
Here, $I_z$ is the moment of inertia of the asteroid, $T_{\varepsilon\,\mathrm{NYORP}}$ is the obliquity component of the NYORP torque, whereas $T_{z,\mathrm{NYORP}}$ and $T_{z,\mathrm{TYORP}}$ are the axial components of NYORP and TYORP respectively.

\subsection{TYORP}
For TYORP, we use the approximation, derived by \cite{golubov2019}:
\begin{equation}
\label{tyorp}
T_{z\,\mathrm{TYORP}}=\frac{\Phi R^3}{c}C_T\exp\left(-\frac{\left(\ln{\theta}-\ln{\theta_0}\right)^2}{\nu^2}\right)\left(\cos^2\varepsilon+1\right),
\end{equation}
with $\Phi$ being the radiation constant at the asteroid's orbit, $R$ the mean radius of the asteroid, and $c$ the speed of light. The exponent comes from the analytic theory of TYORP by \cite{golubov2017}, and the term $\cos^2\varepsilon+1$ is taken from \cite{sevecek2016} to account for the obliquity dependence of TYORP. The thermal parameter $\theta$ is defined as
\begin{equation}
\label{theta}
\theta=\frac{(C\rho\kappa\omega)^{1/2}}{((1-A)\Phi)^{3/4}(\epsilon\sigma)^{1/4}},
\end{equation}
Here, $C$ is the heat capacity of the material that produces TYORP, $\rho$ is its density, $\kappa$ thermal conductivity, $\epsilon$ thermal emissivity, $\sigma$ Stefan--Boltzmann constant, and $A$ is the albedo. The latter is defined as the fraction of incident light that is scattered by the surface, and it is assumed to be constant independently on the incidence angle.

For the parameters, \cite{golubov2017} finds for spherical boulders $\nu=1.518$ and $\ln\theta_0=0.580$. $C_T$ depends on the roughness of the surface, and can be different for different asteroids, with the value for asteroid 25143 Itokawa estimated at $C_T=0.0008\pm 0.0005$.

\subsection{NYORP}
The normal YORP can be approximately described by the following equations \citep{golubov2019}:
\begin{eqnarray}
\label{nyorp_z}
T_{z\,\mathrm{NYORP}}=\frac{\Phi R^3}{c}C_z\left(\cos 2\varepsilon+\frac{1}{3}\right),\\
\label{nyorp_eps}
T_{z\,\mathrm{NYORP}}=\frac{\Phi R^3}{c}C_\varepsilon\sin 2\varepsilon.
\end{eqnarray}
For $\sim 50\%$ of asteroids, these equations describe NYORP with a high precision (see \cite{golubov2019} for a more dedicated discussion), for $\sim 30\%$ the precision of the approximation is poor but the qualitative behavior is correct, whereas for $\sim 20\%$ of asteroids the behavior is qualitatively different, with more roots than Eqs. (\ref{nyorp_z}) and (\ref{nyorp_eps}) predict. Generally, the approximations work worse for asteroids with smaller NYORP.

\begin{figure}[hpt]
\includegraphics[width=1.0 \hsize]{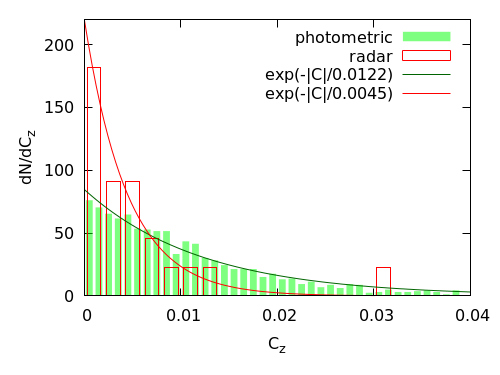}
\caption{\label{C-distr} Normalized distribution over the absolute value of the NYORP coefficient $C_z$ for the photometric shape models from the DAMIT database \citep{damit} and the radar shape models from the JPL Asteroid Radar Research website \citep{radar}.}
\end{figure}

If the thermal inertia of the surface is ignored, the two NYORP coefficients are connected by the equation $C_\varepsilon\approx 0.72 C_z$ \citep{golubov2019}. The distribution of asteroids over $C_z$ is well described by the exponential law (see Fig. \ref{C-distr})
\begin{equation}
\frac{dN}{dC_z}\propto \exp{\left(-\frac{|C_z|}{C_{z0}}\right)}.
\label{B-distribution}
\end{equation}
The constant is estimated at $C_{z0}=0.0122$ for photometric shape models (of mostly Main Belt asteroids) and $C_{z0}=0.0045$ for radar shape models (of Near-Earth asteroids). The difference between these two estimates can either represent the real difference between the two populations, or be attributed to the discrepancy between photometric and radar techniques of shape determination. 

\subsection{Qualitative behavior}
If TYORP and thermal inertia are ignored, asteroids move on smooth trajectories in the $\omega$--$\varepsilon$ space where  $\varepsilon$ changes monotonically, while $\omega$ first increases and (if the asteroid is not disrupted because of centrifugal forces) then decreases. Whether $\varepsilon$ grows or decreases is determined by the sign of $C_\varepsilon$. At the end of evolution, $\omega$ tends to 0 (tumbling), and $\varepsilon$ tends to either 0 or $90^o$ with equal probabilities.

An additional complication arises from the possibility of YORP equilibria due to TYORP compensating NYORP \citep{golubov2019}. These stable equilibria can serve as sinks for the YORP evolution, from which asteroids can be kicked out only by collisions or by change of their orbits.

Consideration of the thermal inertia of the asteroid can complicate the behavior and produce more types of equilibria \citep{scheeres08}.

Still one more complication sets in if we consider the $\sim 20\%$ of asteroids for which the behavior of NYORP is qualitatively different from Eqs. (\ref{nyorp_z})-(\ref{nyorp_eps}). We do not expect these asteroids to significantly effect the results of our simulations, as these are the asteroids with the smallest NYORP, and the dominant contribution to their dynamics is provided by other factors, such as TYORP and collisions.

\section{The statistical model}
\label{MODEL}

Starting from the above theoretical considerations, we have developed a numerical model that computes the dispersion in semi--major axis of a population of asteroids due to the YORP--Yarkovsky coupling. The evolution in time is computed through a series of discrete time--steps during which the  obliquity $\varepsilon$ and the rotation rate $\omega$ of each body are updated  because of the changes due to NYORP-TYORP and collisions. The semi--major axis variation due to the Yarkovsky effect is then calculated at the end of the time--step from the updated spin axis. 

\subsection{Numerical implementation of YORP}

The differential equations for both $\omega$ and $\varepsilon$ are solved using a simple leapfrog method with a time--step much shorter than the evolution period. However, while the analytical solutions are continuous, in the evolution of a real asteroid there there are two sources of discontinuity. The first is when the rotation rate becomes very small and the asteroid enters a  temporary tumbling state  before starting a new YORP cycle, the second is when it rotates fast enough to reach the breakup point. In both cases the analytical solution must be reset by selecting new initial values of $\omega$ and $\epsilon$  and, in the case of breakup,  by drawing new constant coefficients $C_z, C_e, n_0$ for NYORP and TYORP, respectively.

 The way in which we deal with the two singularities, breakup and very slow rotation tumbling state, is derived from the NYORP evolution curves shown in  Fig.\ref{nyorp-coll}.
We first define the minimum rotation rate $\omega_{tum}$ setting it to $10^{-3} \quad rev/day$ (period equal to 1000 days). 
The choice of this value is rather arbitrary and it must be taken into account that close to this limit, also a collision with a small projectile can significantly change  the rotation rate and move the body on a different NYORP track. 
When the threshold value $\omega_{tum}$ is reached, for obliquity values $\epsilon \sim 0, 90^o, 180^o$, we 
change the obliquity by a small amount so that the body evolves out of tumbling slowly accelerating its spin rate and starting a new NYORP cycle towards faster rotation rates. Since the NYORP paths are traveled from left to right, for continuity when the slow rotation tumbling state is reached close to $\epsilon = 90^o$, the body is taken out of the tumbling state by slightly increasing the obliquity so that $\epsilon > 90^o$ and the NYORP cycle evolves towards faster rotation rates. If instead the slow rotation is achieved close to $180^o$, then we select for the obliquity a small value $\epsilon \sim 0$ and a new NYORP cycle is started. In all cases, the rotation rate $\omega$ is reset to $\omega_{tum}$.

To deal with the breakup limit,  we set a threshold value of   $\omega_{dis} = 9.6 \quad rev/day$ corresponding to a period of 2.5 hr. This is slightly larger than the  critical disruption spin rate amongst the asteroid population predicted by \cite{pravec2007} ($\sim 2.3 $ hr) since the reshaping may begin earlier on \citep{walsh2008} depending on the body internal structure. In addition, this limit appears to depend on the shape and density of asteroids and their taxonomic type \citep{pravec2000,chang2015} with values larger than 3 hrs for low density C--type asteroids. In this scenario, our choice appears conservative and can be refined once a specific family will be considered. 

When  $\omega_{dis}$ is reached during a YORP cycle, we expect that the shape and the surface features of the asteroid are altered due to the development of landslides. Therefore, new coefficients $C_z, C_e, n_0$ are drawn and, at the same time, a new value of $\epsilon$ is randomly selected in between $30^o$ and $90^o$ or $120^o$ and $180^o$. 
By inspecting Fig.\ref{nyorp-coll}, we observe that the peaks of the NYORP cycles are located within these intervals.  We could have chosen wider ranges for the new values of $\epsilon$, but our choice is more conservative by avoiding extreme NYORP cycles. The new value of $\omega$ with which the body evolves to slower rotation rates is reset to $\omega_{dis}$. A few tests with different values of both $\omega_{tum}$, which has been changed by a factor 10, and $\omega_{dis}$ show that the overall evolution does not significantly depend on the choice of these parameters. 

\subsection{Calculation of the Yarkovsky drift}

To estimate the value of the semi--major axis drift of the family members, we first numerically compute the dynamical evolution of some of them with the symplectic integration code SWIFT-RMVS3 modified to accommodate Yarkovsky thermal forces \citep{bottke2001,scholl2005} setting reasonable values for the parameters of the Yarkovsky force. We select two initial values of $\epsilon$ i.e. $\epsilon = 0^o$ and $\epsilon = 90^o$, an albedo of 0.3,  a bulk density of $2.5 \quad g/cm^3$, a surface density of $1.5 \quad g/cm^3$, a surface conductivity $K = 0.001 \quad W/(m K)$, and an emissivity of 0.9. We then scale the numerically computed $da / dt$ with $\omega$, following \cite{farnocchia2013}, and with $\cos(\epsilon)$ for the diurnal component and with $\sin^2(\epsilon)$ for the seasonal one.  This approach gives a reasonable value of the Yarkovsky drift rate and, even if with some approximation, shows how the family members evolve differently under the variations of both $\omega$ and $\epsilon$ due to YORP. 

\subsection{Collisions}  The change in the rotation state of an asteroid due to repeated collisions with other Main Belt asteroids is modeled as in \citep{marzari2011Icar}. In short, the population of  potential impactors, derived from \citep{izevic2001}, is divided in logarithmic discrete bins in radius and during each time--step a number of collisions with the potential projectiles in each bin is computed. The intrinsic probability of collision is used to compute the frequency of collisions within each bin while an impact speed is sampled from the distribution of the impact velocities in the asteroid belt  as derived in \citep{farinella92, bottke1994, vedder1996, vedder1998}.  The Poisson statistics is then used to compute for each family member a list of collisions characterized by the time of impact, the size of the projectile and the relative velocity.  For each impact on the list we compute a collision geometry (alt--azimuth angles and impact parameter), randomly defined within the limits given by the orbital element distribution of asteroids in the Belt. The  angular momentum of the projectile is vectorially added up to that of the target and the rotation rate and obliquity of the target are updated. If the impact energy is high enough for fragmentation,   we assume that the target is shattered  and draw a new object from the initial distribution.
We neglect the angular momentum taken away by the fragments that may escape after the cratering. This approximation is good for the frequent low energy impacts, but is less accurate for the few very
energetic impact events. However, as stated above, these events
are not really important for the overall evolution of the spin rate. The main effect of collisions is to cause a random walk of the angular momentum and to change the strength of both NYORP and TYORP, which depend
on the shape and surface features of the asteroid. In fact, anytime a collision occurs, if it is very energetic we update both the $C_z, C_e$ and $n_0$ of the family member while for the less violent impacts we update only $n_0$. 
The choice of the energy threshold for the change in the YORP parameters is somehow arbitrary since we do not have at present precise predictions on the change of the coefficients as a function of the impact energy. We expect TYORP to be more sensitive to the formation of craters on the surface of the asteroid, so we set a lower limit in the impact energy for the change in the $n_0$ parameter. 

\begin{figure}[hpt]
\includegraphics[width=1.0 \hsize]{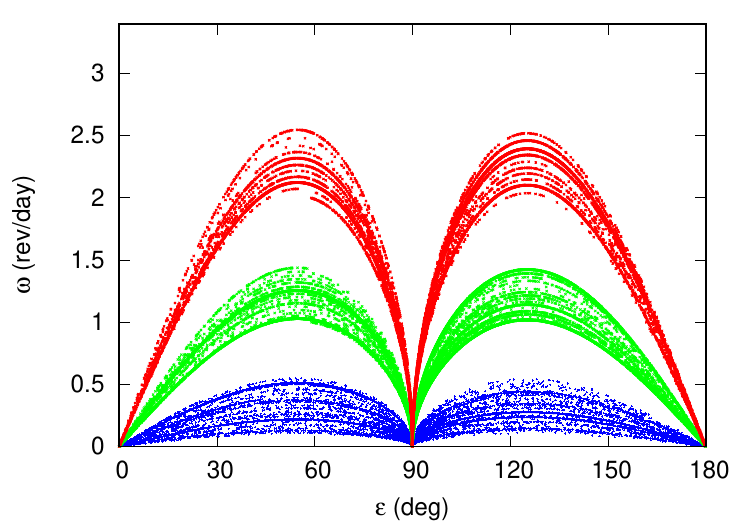}
\includegraphics[width=1.0 \hsize]{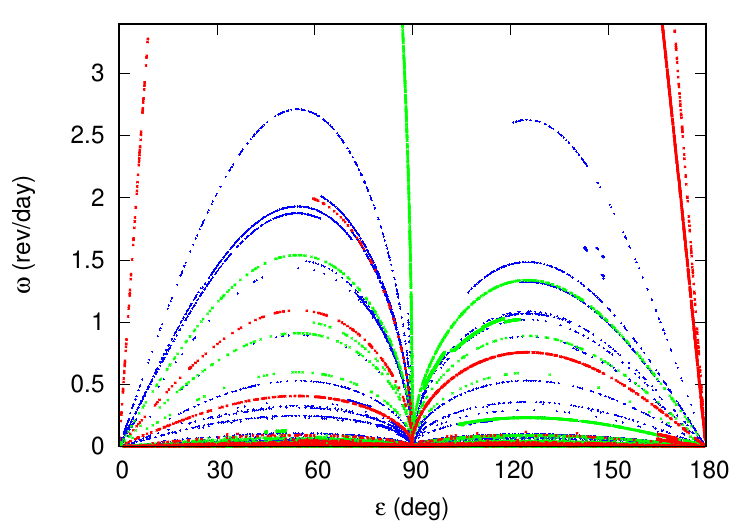}
\caption{\label{nyorp-coll} Effects of collisions on the NYORP cycles. On the top panel only the NYORP effect is included in the spin evolution of three test small asteroids ($D = 2$  km) started with the same spin orientation but different initial rotation rates. On the bottom panel, the same asteroids are evolved including the collisions. We cut the plots at $\omega = 21$ 1/day to better outline the evolution due to NYORP.}
\end{figure}

\begin{figure}[hpt]
\includegraphics[width=1.0 \hsize]{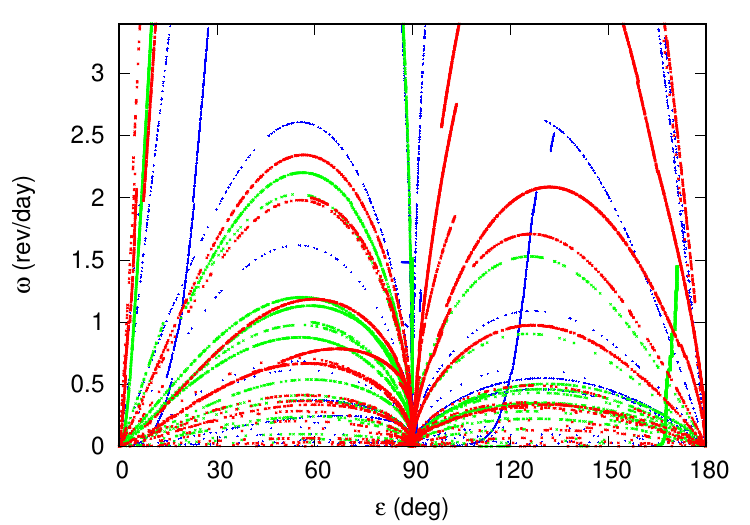}
\caption{\label{tyorp-coll} The same cases shown  in Fig.\ref{nyorp-coll} are now evolved including the TYORP torque  with a coefficient $n_0$ randomly chosen between 0 and 0.02 and collisions. }
\end{figure}

The effects of collisions on the YORP evolution of asteroids is relevant since each impact can change the values of $\omega$ and $\epsilon$ and the more energetic ones also alter the values of the coefficients of the YORP cycles. In Fig.\ref{nyorp-coll} we show the different evolution of the obliquity and rotation rate of three sample asteroids with a small diameter $D=2 \quad km$ all having the same initial direction of the spin axis but different rotation rates ($\omega_1 = 0.1 \quad rev/day$,  $\omega = 1.0 \quad rev/day$,  $\omega = 0.5 \quad rev/day$).  If the effects of TYORP and collisions are neglected (top panel), the body follows smooth NYORP cycles and small changes occur when the body enters a temporary tumbling state at slow rotations. The temporal evolution is from left (exiting the low rotation tumbling state at $0^o$) to right and, again, from left to right for $\epsilon > 90^o$ exiting the slow rotation tumbling state at $\epsilon = 90^o$. In the code, the rotation rate is set to a value which is slightly lower than the threshold limit while reversing the NYORP cycle and for this reason small changes develop within the same cycle. 

When collisions are included (bottom panel), the evolution is more chaotic and the NYORP cycles are almost uncorrelated due to the sudden collisional resetting. Even if the bodies have initially slow rotation rates, they can be driven close to the rotational breakup because of the random walk in the NYORP parameters and the consequent evolution along different NYORP brances, some of which drive to breakup. If also the TYORP effect is included, its tendency to further increase the rotation rate complicates the evolution and more bodies are accelerated towards the breakup limit. This is illustrated in Fig.\ref{tyorp-coll} where the excursions of the spin rate towards the breakup limit are more frequent than in the previous case with collisions only. In conclusion, 
TYORP and collisions together conspire against regular NYORP cycles often driving the bodies towards breakup where the shape and, consequently, the NYORP (and TYORP)  parameters are changed. The behaviour given in the top panel of Fig.\ref{nyorp-coll} is then only speculative and the real evolution is expected to be that described in Fig.\ref{tyorp-coll} even with the due uncertainties in the value of the TYORP parameter $n_0$. 

\section{Test family evolution}
\label{TEST}

To investigate the effects of the YORP evolution and collisions on the 
dynamical spreading of family members due to the
size-dependent Yarkovsky effect, we first generated a simple initial family made of 200 members. The initial semi--major axes are randomly selected between  2.68 and 2.72 au and their diameters are distributed according to a power law with $N(D) = N_0 D^{-3/2}$ and cut at diameters larger than 30 km.  The initial structure of this putative family is over--simplified to avoid features which may be due to the physics of the initial breakup event. In this first study  we want first to understand the dynamical evolution of a cluster of bodies only under the coupling of  YORP, Yarkovsky and collisions. 

\begin{figure}[hpt]
\includegraphics[width=1.0\hsize]{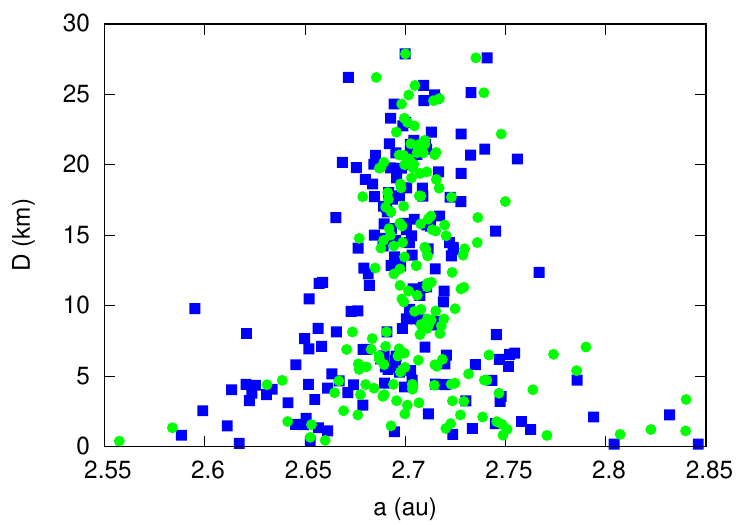}
\caption{\label{fam_noTYORP} Distribution of semi--major axis $a$ and diameter $D$ of the family after 4.5 Gyr. The green filled circles illustrate the case without collisions while the  blue filled squares show the evolution of the family members when the collisions are taken into account. Only the diurnal Yarkovsky effect is considered and no TYORP}
\end{figure}

We first focus on the effects of NYORP and collisions and, for this reason, we neglect the seasonal Yarkovsky effect and TYORP. We performed two simulations where the collisions are included only in the second case and we compare the distribution of the two synthetic families in the $a$ vs. $D$ plane. In Fig.\ref{fam_noTYORP} shows the final distributions of the 
family members after 4.5 Gyr of evolution. 
A higher dispersion in semi--major axis for diameters larger than 5 km  is observed when the collisions are included (blue filled squares) compared to the case without collisions (green filled squares), and this is possibly related to the loss of coherence in the obliquity evolution. 

\begin{figure}[hpt]
\includegraphics[width=1.0 \hsize]{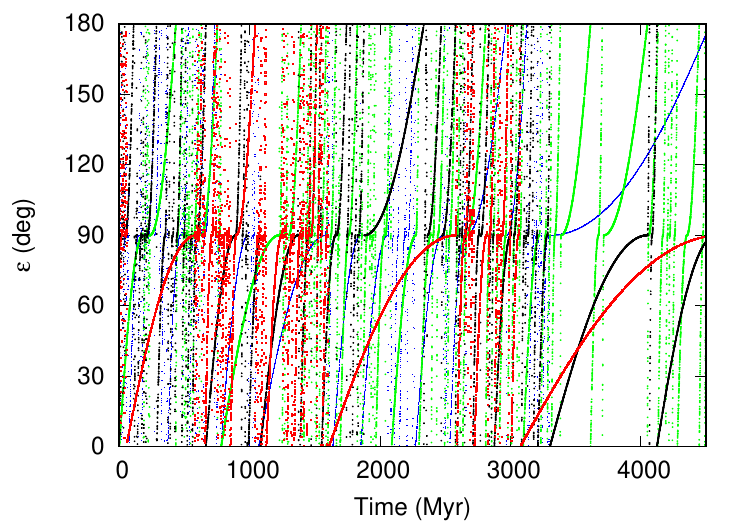}
\includegraphics[width=1.0 \hsize]{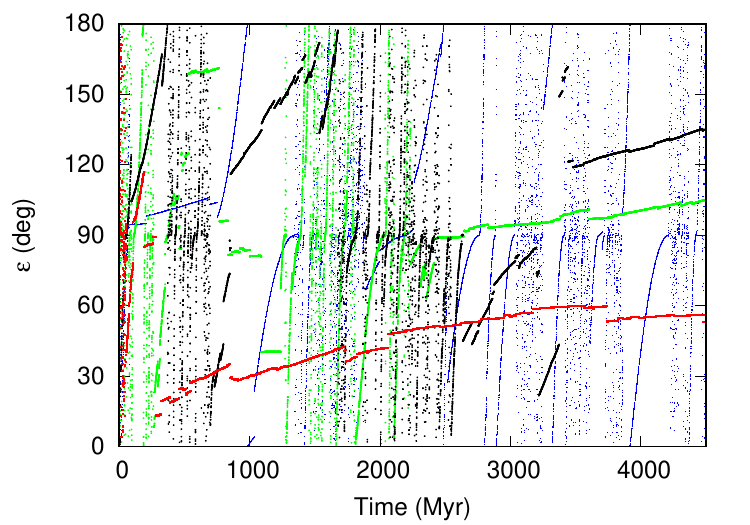}
\caption{\label{fam_dispersion}
Evolution of the obliquity $\epsilon$ for 4 putative family members. In the top panel only NYORP is considered while in the bottom panel collisions are included in the simulation.}
\end{figure}

\begin{figure}[hpt]
\includegraphics[width=1.0\hsize]{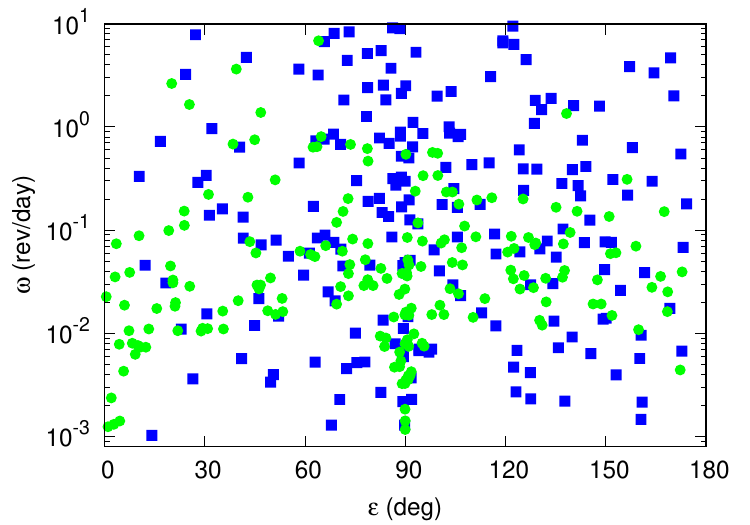}
\caption{\label{eps_distri}
Final distribution of the rotation properties of the family members in the case without (green filled circles) and and with collisions (blue filled squares).}
\end{figure}

Another interesting aspect of the family evolution is related to the distribution of the spin of the family members. 
In  Fig.\ref{fam_dispersion} we show the time evolution of the obliquity of four test bodies  of similar size  $D=2 km$. In the case without collisions (top panel) regular YORP cycles are observed with different periods, depending on $C_z$. There is an overdensity of values around $90^o$ where the bodies slowly approach and exit from the tumbling state. This is further confirmed in Fig.\ref{eps_distri} showing the distribution of $\omega$ vs $\epsilon$ at the end of the simulation with 200 family members. The green filled squares, showing the final values of $\omega$ and $\epsilon$ in the model without collisions, are concentrated around $90^o$ for slow rotation rates. This clustering is related to the long time required to reach the slow rotation tumbling state when the obliquity approaches $90^o$ and to exit from it evolving towards $180^o$. Note that the curves in Fig.\ref{nyorp-coll} and  Fig.\ref{tyorp-coll}  travel from left to right between $0^o$ and $180^o$ with a singular point at $90^o$. This concentration around  $90^o$ might
favor the seasonal Yarkovsky (which is not included in these runs)  over the diurnal one even if the semi--major axis drift is expected to be slower due to the slower rotation rates of the bodies 
\citep{farnocchia2013}.

In the case with collisions (bottom panel of Fig.\ref{fam_dispersion}), the evolution of the four test bodies is irregular due to the changes in the NYORP parameters which not only occur at breakup but also after energetic collisions. This less smooth behaviour weakens the concentration of the obliquity around $90^o$ and the distribution in the $\varepsilon$ vs. $\omega$ plane appears more randomized  as illustrated in Fig.\ref{eps_distri} by the blue filled squares. 

\begin{figure}[hpt]
\includegraphics[width=1.0\hsize]{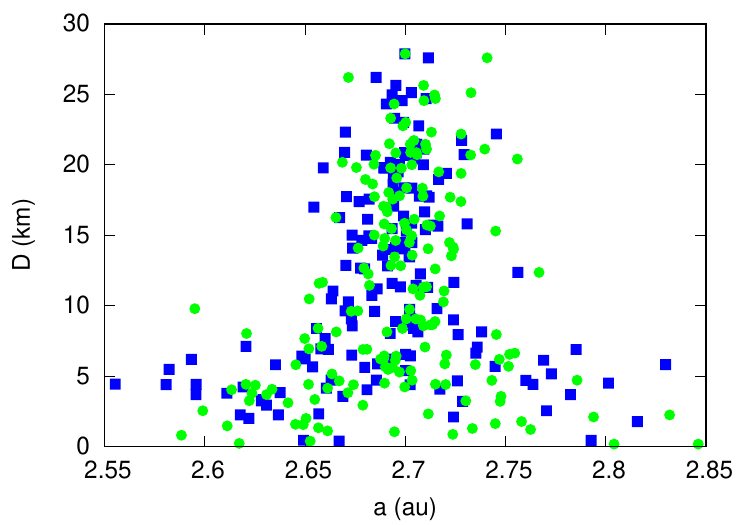}
\caption{\label{TYORP_NYORP}
Same as in Fig. \ref{fam_noTYORP} but in this case the blue filled squares represent the synthetic family distribution after 4.5 Gyr with NYORP, TYORP and collisions. The green filled circles are instead the outcome of a simulation without TYORP (only NYORP and collisions).}
\label{T-NYORP_COLL_family}
\end{figure}

We finally add TYORP in the model and in Fig. \ref{TYORP_NYORP} we compare the final semi--major axis distribution with and without TYORP (collisions are included in both cases). The contribution from TYORP does not seem to increase the family spreading with respect to the case with only NYORP. It means that TYORP significantly affects the evolution of the individual spin of a body but, on average, it does not affect the final distrubution of the family in the $a$ vs $D$ plane.  In this  model the TYORP coefficent $n_0$, which is proportional to the number of boulders on the
surface, is randomly selected in the range (0,0.02). For asteroid (25143) Itokawa,  where numerous boulders have been 
identified on its surface,  a value of 0.03 has been estimated by \cite{sevecek2015} but even higher values might be possible depending on the evolutionary history of the body. However, it is difficult to estimate the density and shape of boulders on main belt asteroids and for this reason we adopt a statistical approach where we select a random value for each body. The upper limit assumed in the simulation is $n_0 = 0.02$, somewhat arbitrary and smaller than that of Itokawa. It is based on the assumption that even monoliths are present in the belt with a potentially lower boulder density on their surface. At present it is difficult to achieve a more reliable estimate.

\begin{figure}[hpt]
\includegraphics[width=1.0 \hsize]{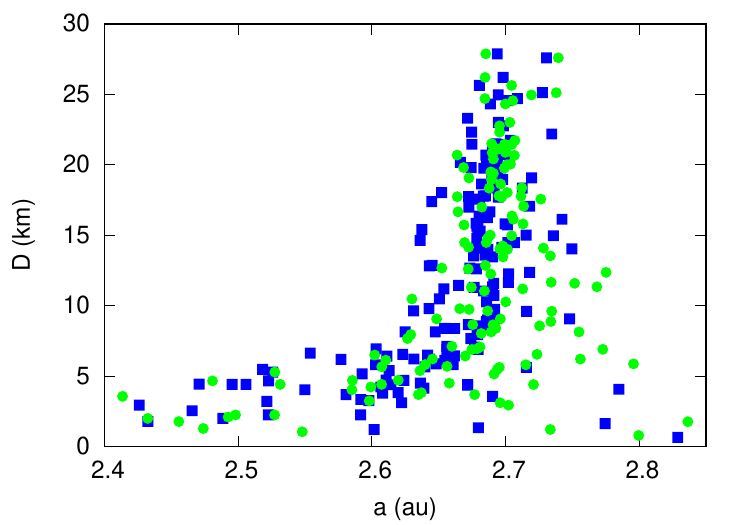}
\includegraphics[width=1.0 \hsize]{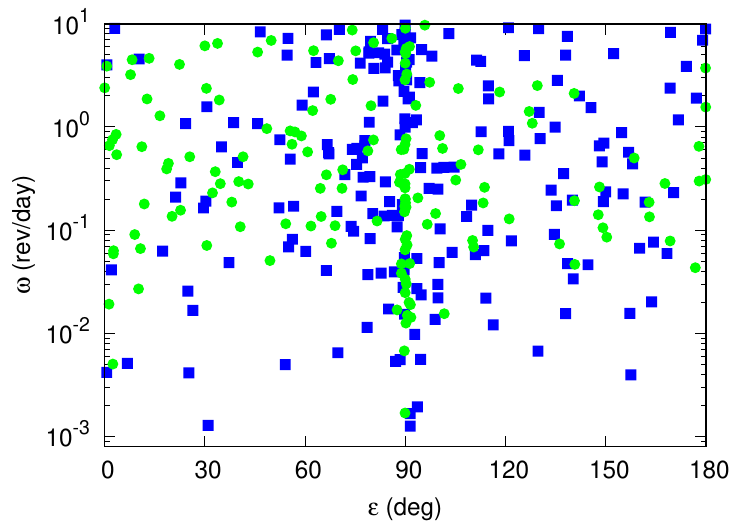}
\includegraphics[width=1.0 \hsize]{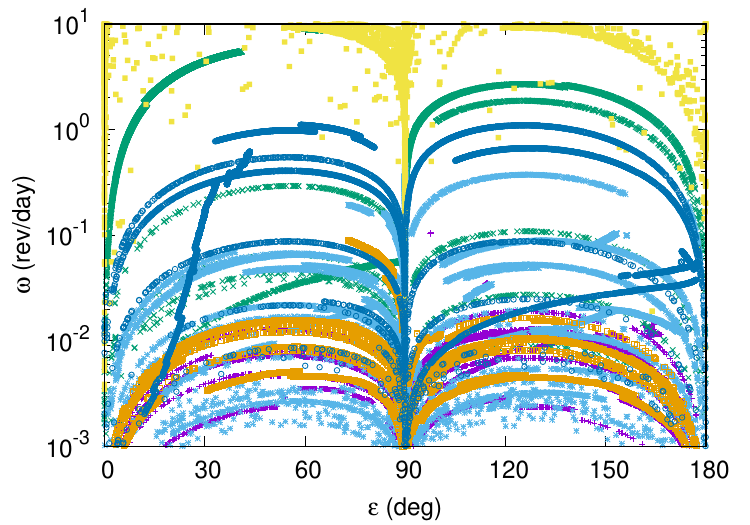}
\caption{\label{seasonal}
Same as in Fig.\ref{fam_noTYORP} but with the seasonal Yarkovsky effect included in the numerical model. We consider two different values of $n_0$ for the TYORP effect, $n_0=0.02$ (blue filled squares) and $n_0 = 0.005$ (green filled circles). In the top panel we show the final semi--major axis distribution of the putative family after 4.5 Gyr, in the middle panel the final distribution of  $\omega$ vs. $\epsilon$  while in the bottom panel the time evolution of  $\omega$  vs. $\epsilon$ for some sample bodies in the strong TYORP case. }
\end{figure}

After the testing with only the diurnal Yarkovsky effect, leading to a mostly symmetrical distribution of the putative family members around their initial semi--major axis, we also included the seasonal term in the model  (Fig.\ref{seasonal}). The family is shifted inwards, as expected, with some small members drifting very far from the initial location. For these bodies the obliquity is lingering close to $90^o$ for an extended period of time.  
Two different values have been tested for TYORP,  $n_0 = 0.02$  (green squares) and $n_0 = 0.005$ (blue squares). While for the semi--major axis there are not significant differences in the two cases (top panel),  when TYORP is weaker there are fewer bodies with slow rotation rates and $\epsilon$ located close to either 0 or $180^o$. In the bottom panel of the figure the evolution of a small sample of family members  is illustrated as typical example of the $\omega$ vs. $\epsilon$ evolution with time. In all these cases $n_0 = 0.02$ is adopted. 

\section{Test on a large Koronis--like  family} 
\label{TEST_BIG}

\begin{figure}[hpt]

\includegraphics[width=\hsize]{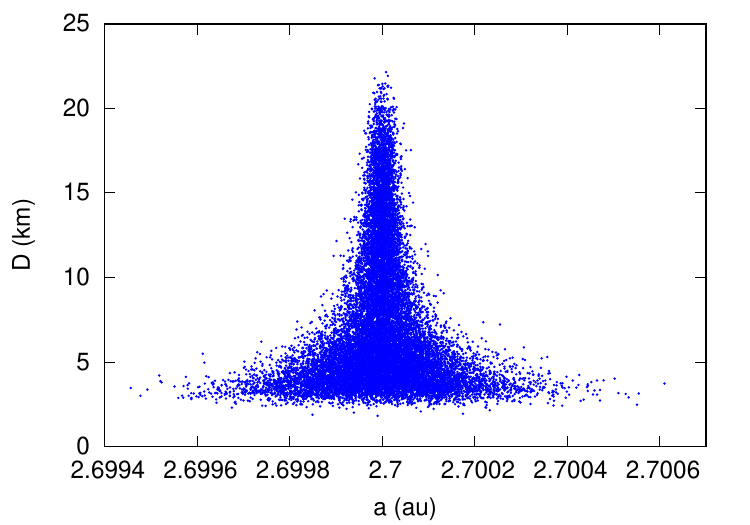}
\caption{\label{aldo_initial} Initial distribution  of  semi–major  axis
$a$ and diameter $D$ for the a large family. Note the smaller scale in semi--major axis compared to the evolved family.}
\end{figure}

A further test was completed on a simulated family with about 20000 members in order to have a richer statistics.  The family members were generated with a power law having an exponent such that the distributions of their diameters ($D$) is uniform in a $\log D$ scale. An ejection velocity vector $V$ and a spin vector $S$ are assigned at each member in the following way. The direction of $V$ is chosen at random, assuming an isotropic symmetry of the ejection velocity field. The modulus $V$ is randomly generated according to a Maxwellian distribution, the mean value of which is related to the diameter. If $D_0$ is a reference diameter (usually 1 km), the mean value of the distribution of $V$ is $V_m(D) = V_0 (D_0 /D)^\beta$ , where $\beta = 2/3$. Also the direction of $S$ is assumed isotropically distributed. Its modulus $S = 2\pi/P$, where $P$ is the rotation period of the asteroid, is generated according to another Maxwellian distribution, the mean value of which
is $S_m(D) = S_0 (D_0 /D)^\gamma$ , where $\gamma=5$.  The values
of the exponents $\beta$  and $\gamma$ 
have been selected to satisfy the equipartition of total kinetic energy and   total angular momentum
among all the members of the family \citep{cellino1999}. 
With this choice, the velocities and spin rates tend to increase as the size of the bodies decreases, at a pace depending on the values of the exponents $\beta$ and $\gamma$. 
Finally the initial orbital elements of the members were computed from the corresponding ejection velocities and the orbital element of the parent body at the moment of the break-up. In general, the resulting distribution of the members' semi-major axes depends mainly on the semi-major axis of the parent body and the transversal components of the ejection velocities, while it is much less affected by the other orbital elements, in particular the true anomaly of the parent body. The family parent body is located at 2.7 au. Note that for the current tests we are considering objects between 5 and 50 km of diameter. In Fig~\ref{aldo_initial} we plot the initial distribution of the family members in the semi--major axis vs. size plane. 

\begin{figure}[hpt]

\includegraphics[width=\hsize]{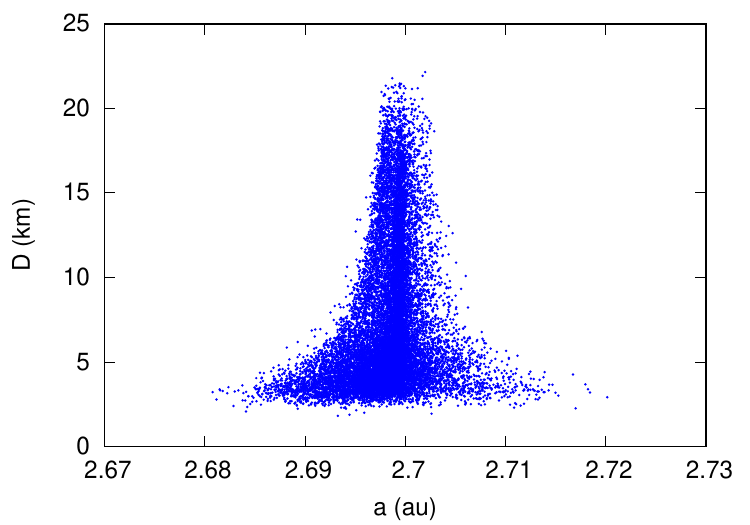}
\includegraphics[width=\hsize]{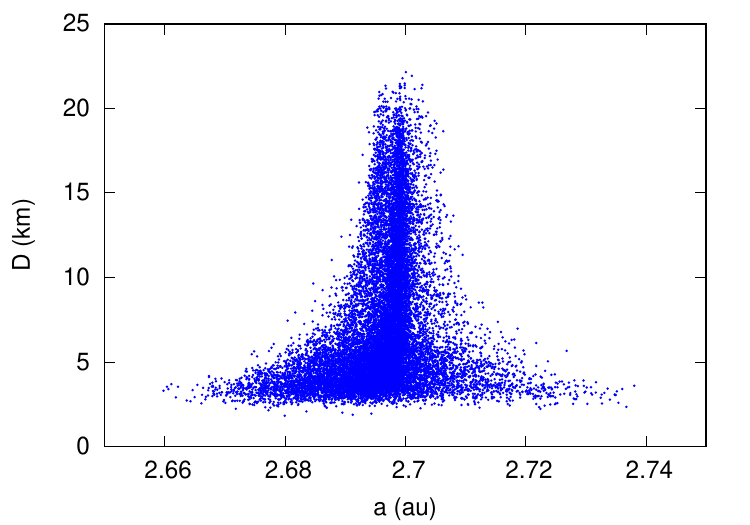}
\includegraphics[width=\hsize]{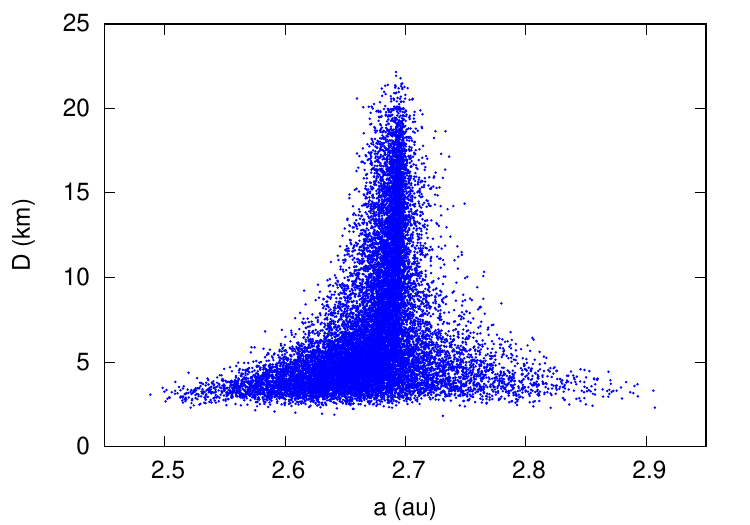}
\caption{\label{fam_time} Distribution of   semi–major  axis
$a$ and diameter $D$ for the large family at different evolutionary times. The top panel shows the family after 250 Myr, the middle panel after 500 Myr and the bottom panel after 2500 Myr. Note the progressive expansion of the semi--major axis scale due to the increasing spread of the family with time. }
\end{figure}

\begin{figure}[hpt]
\includegraphics[width=\hsize]{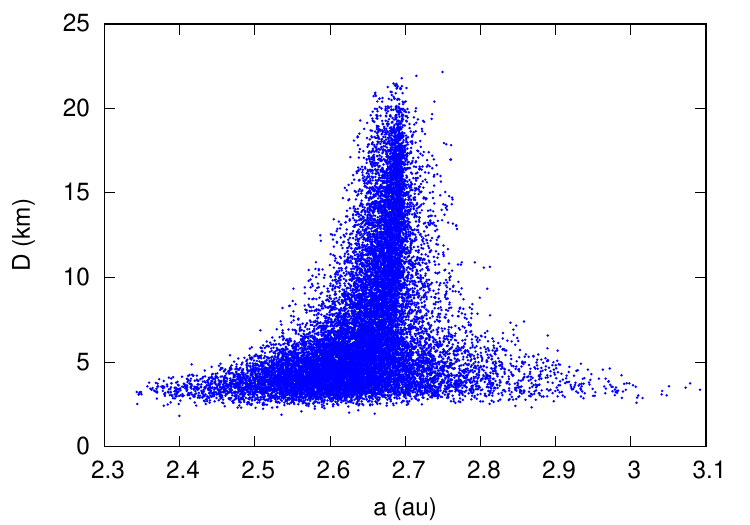}
\includegraphics[width=\hsize,angle=0]{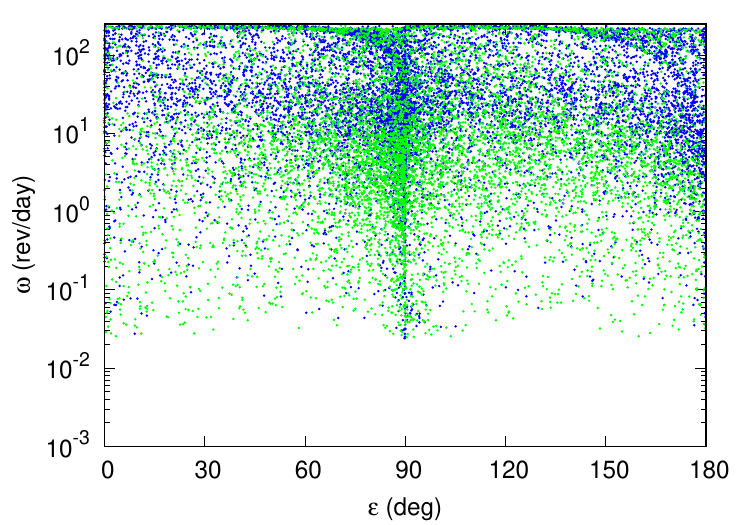} 
\caption{\label{fam_final} Distribution of semi--major axis
$a$ and diameter $D$ of the large family after 4.5 Gyr (top panel). In the bottom panel we show the $\omega$ vs. $\epsilon$ for the same family. The green dots are bodies with size larger than 3 km in diameter while the blue ones are smaller bodies.}
\end{figure}

 The synthetic family is evolved for  4.5 Gyr and in Fig.~\ref{fam_time}
 we show three snapshots at t= 250, 500 and 2500 Myr. A progressive spreading of the family in semi--major axis is observed, due to the Yarkovsky drift, and  some unevenness develops after 500 Myr  possibly related to the initial conditions of the family. However, after 2500 Myr the distribution of the family members appears homogeneous since the  YORP cycles are randomized by collisions. The family is significantly more dispersed towards smaller semi--major axes due to the seasonal Yarkovsky.   This bending, which begins to develop already after 250 Myr and is particularly noticeable after 4.5 Gyr (see Fig.\ref{fam_final}), is reinforced by the obliquity distribution which shows a concentration around $90^o$, as illustrated in the bottom panel of Fig.~\ref{fam_final}.
Despite collisions,  the obliquity values cluster around $\epsilon=90^o$ because of the time evolution during the NYORP cycles. When the obliquity move towards  $\epsilon =90^o$ and $\omega$ approaches the slow rotation tumbling state, the $\omega$ decreasing trend slows down keeping the body close to $\epsilon =90^o$ for an extended period of time. The same happens when the body progressively leaves the slow rotation tumbling state with $\epsilon > 90^o$ towards faster rotation rates. As also illustrated in Fig.\ref{fam_dispersion}, close to $\epsilon =90^o$ the NYORP cycles slow down and this explains the crowding of the family members around $\epsilon =90^o$ and the increased efficiency of the seasonal Yarkovsky effect. 

It is noteworthy that the smaller members of the family (blue dots) have on average faster rotation rates. They are more affected by collisions that tend to increase $\omega$  and, in addition, they interrupt the slow approaching of a tumbling state during a NYORP cycle resetting its evolution.

\section{Conclusions}
\label{CONCLU}

Based on the most recent developments of the YORP theory \citep{golubov2019}, we built a statistical model computing the semi--major axis drift due to the Yarkovsky effect, coupled to YORP, of a putative population of asteroids, like a family, including also the overall effect of collisions on the asteroid rotation rate. The latter can significantly alter the smooth evolution of  both $\varepsilon$ and $\omega$ predicted by YORP not only via impulsive changes in the angular momentum vector but also by resetting the subsequent YORP evolution from then on and changing the YORP parameters.  An energetic cratering collision can in fact change the number of boulders on the surface of the body and probably increase their heat conductivity altering, as a consequence, the strength of TYORP. At the same time, if the impact is violent or in presence of re--shaping due to rotational breakup, also the NYORP parameters change leading to a different spin vector evolution. We checked our algorithm on a small test family switching on and off the different effects like TYORP, collisions and seasonal Yarkovsky.  We have found that collisions force a random walk of the spin vector which significantly departs from the smooth and regular one predicted by NYORP only. The NYORP resetting due to the different values of $\omega$ and $\varepsilon$ after each collision and the changes in the NYORP parameters lead to a more complex evolution of the spin axis that translates into a higher dispersion of the family in semi--major axis also for larger bodies. When TYORP is included, the regularity of the spin axis evolution is further reduced and a higher percentage of bodies reach the breakup.  We also performed an additional test on a significantly larger family to understand how much of the initial family structure may be preserved in time against the coupled  Yarkovsky--YORP evolution and to have a richer statistics of the final values of $\varepsilon$ vs. $\omega$.
 A clustering is observed close to $90^o$ due to the asymmetry in the time evolution of the NYORP cycles  while approaching the slow rotation tumbling state. 

In the future, we intend to model some real families to test
how they evolve in time and how this evolution relates to the YORP model parameters like $C_z$ for NYORP and $n_0$ for TYORP. Once the model is fine tuned on well studied families, it can be used, for example,  to study and predict the structure and age of other families.  
 At this stage we will also distinguish between fragmentation and cratering events which  have different initial conditions for the cluster of fragments and may lead to a different evolution and dynamical structures.   

Finally, we point out that in these test runs we did not observe the formation of a 'YORP--eye' as the one predicted by \citep{paolicchi2016,paolicchi2019} since, according to Fig.\ref{eps_distri}, Fig.\ref{seasonal} and Fig.\ref{aldo_seas_no_seas}, the distribution of the obliquity  peaks around $90^o$ and not at $0^o$ or $180^o$ as required by the 'YORP--eye' formation.  This raises the question whether the 'YORP--eye', which seems to be present in some families like 5124 2000EZ39 or 845 Naema \citep{paolicchi2019}, may be a feature related to the collisional physics rather than to the subsequent evolution driven by the Yarkovsky effect. For this reason in the future  we will perform  simulations with different initial family structures, related to the breakup physics, and test for how long  potential initial features in the $a$ vs. $D$ plane survive and can be detected. 

\section{Acknowledgments}
\label{Ack}
We would like to acknowledge the help and discussions with Paolo Paolicchi and Aldo Dell'Oro on several aspects of this work and, especially, in the definition of the large synthetic family. We also thank two anonymous referees for their helpful comments and suggestions. 

\bibliographystyle{aasjournal}
\bibliography{biblio}


\end{document}